# Transfer of multi-DNA patches by colloidal stamping

Rawan Khalaf,[a†] Andrea Viamonte,[a] Etienne Ducrot,*[a] Rémi Mérindol*[b] and Serge Ravaine*[a]

Patchy particles have received great attention due to their ability to develop directional and selective interactions and serve as building units for the self-assembly of innovative colloidal molecules and crystalline structures. However, synthesizing particles with multiple dissimilar patches is still highly challenging and lacks efficient methods, these building blocks would open paths towards a broader range of ordered materials and their inherent properties. Herein, we describe a new approach to pattern functional DNA patches at the surface of particles, by use of colloidal stamps. DNA inks are transferred only at the contact zones between the target particle and the stamps thanks to selective strand-displacement reactions. The produced DNA-patchy particles are ideal candidates to act as advanced precision/designer building blocks to self-assemble the next generation of colloidal materials.

## Introduction

Over the past decades, scientists have aspired to fabricate functional materials by colloidal self-assembly.[1] Although many beautiful examples of self-assembled colloidal molecules[2] or colloidal crystals[3] from particles with well-defined shapes and composition have been reported so far, colloidal systems cannot be targeted towards most of the sophisticated structures that Nature built. Indeed, the latter require encoding the building units with information to guide their self-assembly by programming their geometry as well as the directionality, valence, range of their pair interactions. Several strategies have been developed to address this challenge, including the attachment of molecules that recognize one another onto the surface of particles.[4] Among the wide range of binding groups that have been employed, synthetic DNA strands have been proven to be very versatile and promising as a tremendous number of orthogonal interactions can be programmed based on the design of nucleotide sequences, giving access to highly specific programmable interactions. DNA-coated particles have thus been extensively employed as building blocks for the self-assembly of clusters with precise symmetries[5] and crystalline lattices.[6] To further control both the valence of the particles and the directionality of the bonds they form with their partners, a number of groups have recently proposed strategies to regioselectively pattern particles with DNA patches.[7] Sleiman et al. successfully transferred DNA motifs from a parent 3D DNA template to gold[8] and polymeric nanoparticles.[9] Two-[10] and three-[11] dimensional DNA origami structures were used as stamping platforms to transfer DNA inks onto gold nanoparticles. In both cases, the printed nanoparticles were released from the frame by a strand displacement reaction.[12]

In order to create micron-sized particles with several dissimilar patches, we translate this strategy by using colloidal particles coated with DNA inks as stamps. The colloidal stamps can assemble with support colloids via DNA hybridization. The injection of eject strands allowed us to transfer the DNA inks at the contact zones between support and stamp particles (Fig. 1a) leading to patchy particles. We also take benefit of packing constraints to control the number of stamp particles that can park around the support ones[5a], which finally defines the number of transferred patches.

## Experimental

### Synthesis of DNA-coated particles

Azidated 3-(trimethoxysilyl)propyl methacrylate (TPM) particles were prepared through the azidation of chlorine groups present at the surface of particles previously synthesized according to the protocol developed by Wang et al.[6e] (see ESI). After synthesis the particles were imaged by TEM and SEM. Fig. S1 shows that they are spherical and monodisperse in size. Their surface is smooth, which has been shown to be required to allow an homogeneous distribution of DNA strands during the former step.[13] In order to functionalize monodisperse polystyrene (PS) particles with azide groups, we followed the protocol developed by Oh et al.,[6d] which relies on the physical entrapment of an azidated PS-b-PEO copolymer (PS-b-PEO-N$_3$). For this, the PS particles are swollen with tetrahydrofuran (THF) to allow the PS block of the copolymer to penetrate in the PS particles (see ESI). After evaporation of the THF, the PS block of the copolymer is physically trapped in the PS particles while the PEO block and the terminal azide group form a brush at the surface, swollen by water and exposed to the surrounding media. Azidated particles were further functionalized with DNA following the protocol described by Wang et al.[6e] that ensures a dense surface coverage of the colloids with DNA (see ESI). The process relies on the strain-promoted azide-alkyne cycloaddition (SPAAC) to graft DNA strands end-functionalized with a dibenzocyclooctyne moiety (DBCO) onto azide functionalized particles.

### Characterization

Transmission electron microscope (TEM) images were taken using a Hitachi H600 microscope operating at an acceleration voltage of 75 kV. The samples for TEM observation were supported on conventional carbon-coated copper grids. Scanning electron microscope (SEM) images were taken using a Hitachi S4500 microscope at an accelerating voltage of 5 kV. Confocal fluorescence microscopy images were taken using a Leica SP2 confocal laser scanning microscope as well as a ZEISS LSM980 equipped with an Airyscan detector.

## Results and discussion

DNA strands A and B (Table S1) have first been grafted onto azidated TPM and PS particles, respectively. The coated particles are referred to as $TPM_A$ and $PS_B$. We then functionalized/inked the stamp $TPM_A$ particles with the $Ink_{565}$ (see Table S1 for details) by adding a large excess of ink to a suspension of particles (Fig. 1b). The ink consists in two hybridized DNA strands, T-X-A* and $X^*_{565}$-B*, the latter being modified with the fluorescent dye Atto565. As the domain A* is complementary to the sequence A at the surface of the TPM particles, the ink sticks on their surface due to the formation of A/A* duplexes. The particles and strands were maintained in a buffer enriched in magnesium ions and at low temperature in order to strengthen the DNA duplexes and prevent strand migration (see ESI). The excess of ink was subsequently washed away by centrifugation/dispersion steps. As $Ink_{565}$ is also complementary to strands B, the $TPM_A$~$Ink_{565}$ stamp particles were mixed with a 40:1 excess of $PS_B$ particles to form preferentially small clusters with only one $TPM_A$ particle at the core and $PS_B$ satellites (Fig. 1c). This ultimately maximizes the number of one-patch PS particles produced in the process. When the assembly is completed, $Y^*_{488}$-B* strands (labelled with Alexa488) are added to hybridize with the remaining B strands available on the $PS_B$ particles and passivate their surface (Fig. 1d). Fig. 2a shows that clusters made of one $TPM_A$ particle surrounded by $PS_B$ particles were obtained. A detailed analysis reveals that the number of $PS_B$ particles in the clusters varies from 1 to 7. Their relative amounts have been determined by statistical analysis performed over 100 clusters (Fig. 2c-i). Some rare clusters (~5 %) formed of one $PS_B$ particle in contact with two $TPM_A$ particles could be observed as well as a large amount of free $PS_A$ particles. Thanks to the density difference between TPM (1.2 g.cm$^{-3}$ [14]) and PS (1.06 g.cm$^{-3}$), we successfully removed most of these free PS particles (Fig. 2b) by sedimentation in a PBS based buffer solution of intermediate density (1.07 g.cm$^{-3}$) prepared by mixing $H_2O$ and $D_2O$ (see ESI). The relative proportions of the different clusters remained unchanged during this purification stage, as indicated in Fig. 2c-i, proving that the clusters are sufficiently robust and do not break during centrifugation. The final step to form patchy particles consists in disassembling the clusters formed by the stamp and support particles leaving the fluorescent part of the ink on the support particle only at the contact point between them. To do so, we injected the Eject$_X$ strand which binds to the toehold T of Ink$_{565}$ and replaces the strand X*$_{565}$-B*. This breaks the duplex X/X* that was holding the stamp and support particles and results in the release of the support particle, which still carries the red fluorescent strands X*$_{565}$-B* at the former contact point with the stamp (Fig. 1e and S2). Fig. 2j and Movie S1 show that PS particles with one red fluorescent patch are mostly obtained, validating the developed strategy.

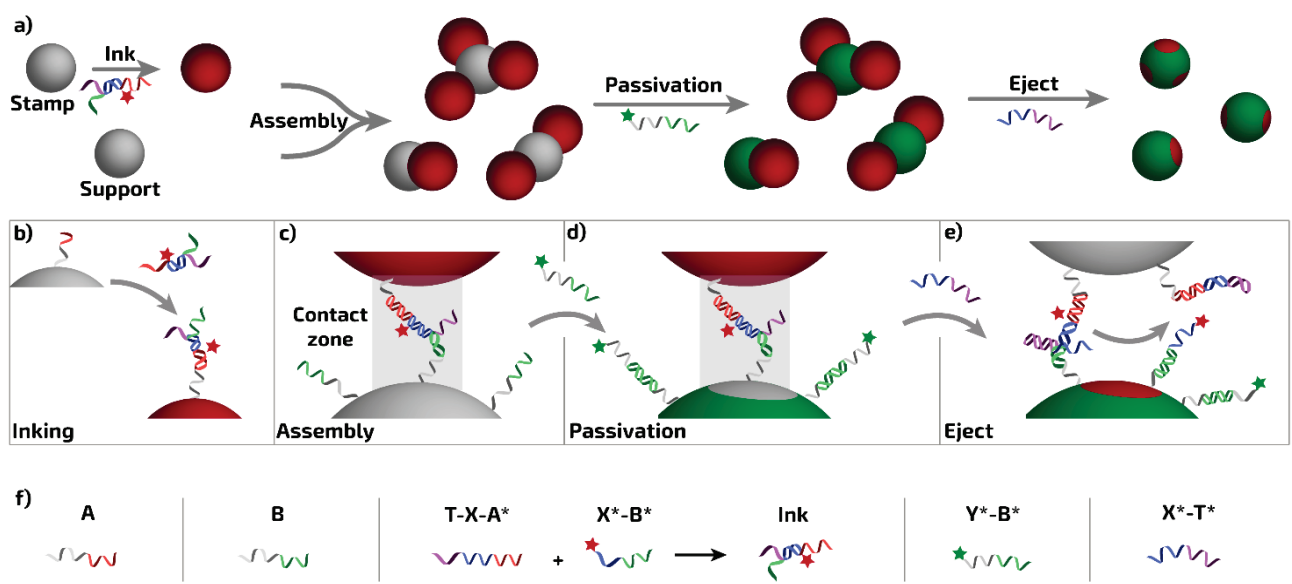

**Fig. 1** General scheme and key steps. a) General schematic representation of the preparation of particles with DNA patches by colloidal stamping following sequentially the subsequent steps. b) Inking: the ink, formed by the association of T-X-A* and X*-B*, is hybridized at the surface of the bare stamp particle decorated with an A* DNA brush. c) Assembly: in the contact zone, formation of duplexes between the support particle decorated by B strands and the stamp thanks to the B* domain exposed by the ink. d) Passivation: the surface of the support particle, outside of the contact zone is passivated by hybridization between the B strands of the surface and Y*-B* strands. e) Eject: strand displacement reaction to separate the stamp from the support particle, leading to the formation of a X*-B* patch at the contact zone and the recovery of patchy particles exposing the stand Y* on the surface with patches of X*. f) Schematic representation of the DNA strands and assemblies used in this study.

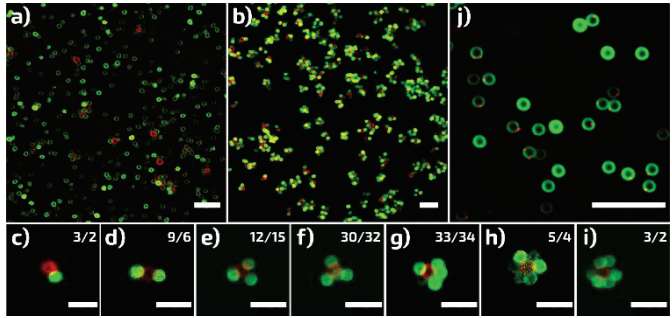

**Fig. 2** Confocal fluorescence microscopy images of clusters obtained by incubating 1.6 μm TPM$_A$~Ink$_{565}$ and 1.5 μm PS$_B$ particles in a 1:40 ratio a) before and b) after purification. c-i) Zoom on the different clusters containing 1 to 7 PS particles. Their relative amounts (in %) in the sample before/after purification are given at the top right corner of each image. j) Confocal fluorescence microscopy image of the patchy PS particles resulting from the strand-displacement disassembly of TPM$_A$~Ink$_{565}$~PS$_B$ clusters. Scale bar of 10 μm for a, b and j and 5 μm for c to i.

Some non-patchy PS particles that were not completely removed by centrifugation and a few two-patch particles (~5 %), which result from the disassembly of the clusters in which one PS particle is in contact with two TPM particles are also observed.

We firstly extended our strategy to prepare particles with multiple identical patches precisely located at their surface. To do so, we prepared clusters with different controlled morphologies by the random parking[5a] of an excess of large PS spheres functionalized with DNA strands A and Ink$_{565}$ on smaller TPM particles functionalized with DNA strands B. Due to packing constraints inherent in the ratio of radii between large and small spheres, only a fixed number of large spheres can park, leading to a population of clusters with well-defined coordination. When the assembly is completed, the Y*$_{488}$-B* strands, complementary to DNA strands B, are added to passivate the surface of TPM$_B$ particles. Fig. 3a-c shows confocal images of the DNA-colloidal clusters obtained when TPM particles with a diameter of 1.1, 1.6, and 2 μm are employed, respectively. Different clusters made of one TPM core and different numbers of PS satellites are observed (Fig. 3d-g and S3). The relative proportions of each type of clusters obtained for different values of the size ratio $\alpha$ of PS/TPM, are listed in Table 1. One can note that when $\alpha$ is 4.39, clusters made of one or two PS particles attached to one TPM sphere are mainly formed. Decreasing $\alpha$ to 3.02 and 2.42 led to the formation of higher proportions of clusters containing 3 and 4 PS particles, as expected.

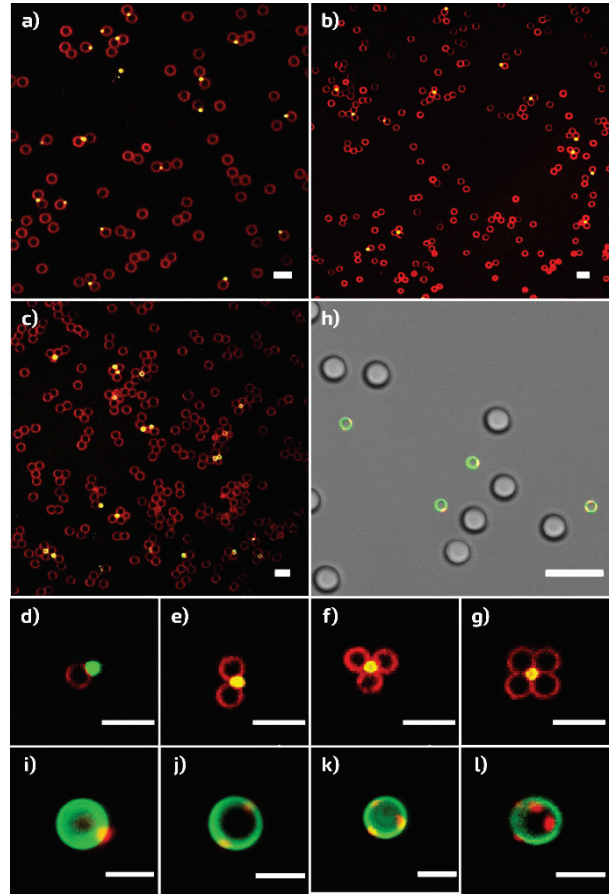

**Fig. 3** Confocal fluorescence microscopy images of the clusters obtained by incubating TPM$_B$ and PS$_A$~Ink$_{565}$ in a 1:40 ratio followed by the addition of the Y*$_{488}$-B* passivation strand. The diameter of the TPM particles is: a) 1.1 μm ($\alpha$ = 4.39), b) and d-g) 1.6 μm ($\alpha$ = 3.02), c) 2 μm ($\alpha$ = 2.42). h) Confocal fluorescence microscopy images (Alexa488, green channel; Atto565, red channel) with transmission microscopy (grey channel) of PS$_A$~Ink$_{565}$~TPM$_B$ & Y*$_{488}$-B* ($\alpha$ = 3.02) clusters after strand displacement reaction using Eject$_X$ strand. i-l) Zoom on the patchy TPM particles with an increasing number of patches obtained after strand displacement reaction using Eject$_X$ ($\alpha$ = 3.02). Scale bar of 10 μm for a to h and 2 μm for i to l.

**Table 1** Compositions of the batches resulting from the mixing of TPMB of different sizes with PSA~Ink565 in a 1:40 number ratio determined by statistical analysis of confocal fluorescence images over about 100 clusters.

| $\alpha$ | ● | ●● | ●●● | ●●●● | ●●●●● |
|---|---|---|---|---|---|
| 4.39 | 3 | 39 | 50 | 8 | 0 |
| 3.02 | 0 | 20 | 29 | 44 | 7 |
| 2.42 | 0 | 3 | 27 | 39 | 31 |

After injection of the eject strands Eject$_X$ in the clusters suspension, non-fluorescent PS particles and TPM particles with red fluorescent patches are observed, proving the transfer of the fluorescent DNA from PS$_A$~Ink$_{565}$ (Fig. 3h and S3). More precisely, Fig. 3i-l show that 1.6 μm TPM particles with one to four patches are obtained. Similar results were obtained with 1.1 μm and 2 μm TPM particles (Fig. S5), validating the strategy based on the combination of colloidal parking and colloidal stamping.

Lastly, we further extended our strategy to prepare particles with multiple dissimilar patches. We first divided the PS particles into two batches, and coated one batch with $Ink_{565}$ and the other with $Ink_{647}$ (Table S1). The two batches were then mixed together and $TPM_B$ particles functionalized with B strands were added in a number ratio $TPM_B:PS_A\sim Ink_{565}:PS_A\sim Ink_{647}$ of 1:20:20. The sample was kept in the fridge at 4 °C for 24 h to maximize the formation of clusters. Then, $Y^*_{488}$-$B^*$ strands were added to hybridize with the B strands outside of the contact zones and passivate the surface of the $TPM_B$ particles.

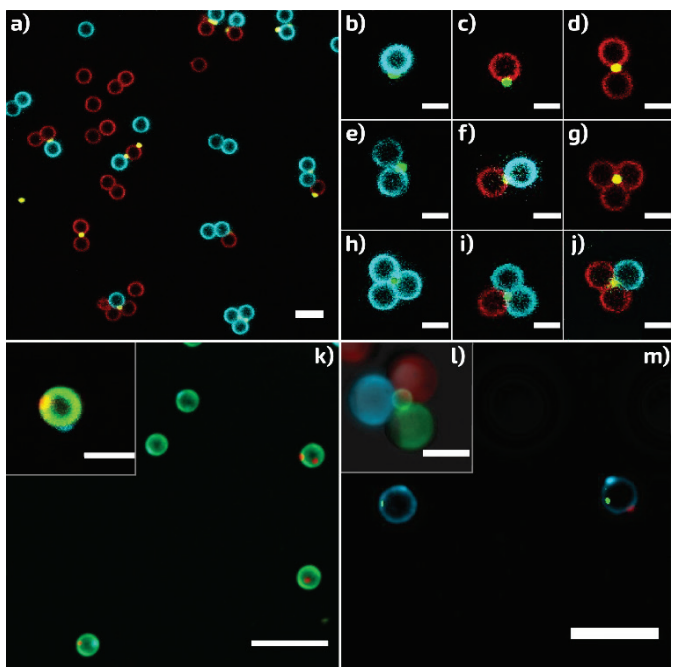

**Fig. 4** Confocal fluorescence microscopy images of the clusters obtained by incubating $TPM_B$, $PS_A\sim Ink_{565}$ and $PS_A\sim Ink_{647}$ in a 1:20:20 number ratio followed by the addition of the passivation strand $Y^*_{488}$-$B^*$. The diameter of the TPM particles is 1.6 μm (α = 3.02). k) Confocal fluorescence microscopy image of the patchy particles obtained after injection of $Eject_X$ and $Eject_Z$. The inset shows a TPM particle with one red and one blue fluorescent patches. l) Confocal fluorescence microscopy image of a cluster made of one TPM particle surrounded by one $PS_A\sim Ink_{488}$, one $PS_A\sim Ink_{565}$ and one $PS_A\sim Ink_{647}$ particles. m) Confocal fluorescence microscopy image of the patchy 1.6 μm TPM particles obtained after injection of $Eject_X$, $Eject_Y$ and $Eject_Z$. Scale bar of 10 μm for a, 5 μm for b to m, and 2 μm for k inset.

Different clusters made of one TPM core and different numbers of $PS_A\sim Ink_{565}$ and $PS_A\sim Ink_{647}$ are observed (Fig. 4a-j). The relative proportions of each type of clusters are listed in Table 2. After injection of the eject strands $Eject_X$ and $Eject_Z$ in the clusters suspension, non-fluorescent PS particles and TPM particles with red and/or blue fluorescent patches are observed, proving the transfer of the fluorescent DNA from $PS_A\sim Ink_{565}$ and $PS_A\sim Ink_{647}$ (Fig. 4k and S6 and Movie S2). When we worked with three batches of $PS_A$ particles coated with $Ink_{488}$, $Ink_{565}$ and $Ink_{647}$, respectively, and mixed them with $TPM_B$ particles in a number ratio $TPM_B:PS_A\sim Ink_{488}:PS_A\sim Ink_{565}:PS_A\sim Ink_{647}$ of 1:13:13:13, we observed the formation of a few clusters made of one TPM particles surrounded by varying numbers of $PS_A\sim Ink_{488}$, $PS_A\sim Ink_{565}$ and $PS_A\sim Ink_{647}$ particles (Fig. 4l). After injection of $Eject_X$, $Eject_Y$ and $Eject_Z$, TPM particles with red and/or blue and/or green fluorescent patches are observed, proving once again the efficiency of our approach (Fig. 4m and Movie S3).

**Table 2** Compositions of the batches resulting from the mixing of $TPM_B$ with $PS_A\sim Ink_{565}$ and $PS_A\sim Ink_{647}$ in a 1:20:20 number ratio determined by statistical analysis of confocal fluorescence images over about 100 clusters.

| 🟢 | 🔴🔴 | 🔵🔵 | 🔴🟢🔵 | 🔵🟢🔵 | 🔴🔵🔵 | 🔴🔴🔴 | 🔵🔵🔵 | 🔴🔵🔵 | 🔴🔴🔵 |
|---|---|---|---|---|---|---|---|---|---|
| 7 | 22 | 26 | 8 | 19 | 13 | 0 | 2 | 2 | 1 |

## Conclusions

In conclusion, we have synthesized micron-sized particles with one or several identical and distinct DNA patches by combining colloidal parking and the transfer of DNA strands at the contact zones with colloidal stamps thanks to strand-displacement reactions. Our strategy is versatile and can be extended to a gallery of hard particles and soft systems whose shape and size could be independently varied, opening the way to the synthesis of new DNA-patchy building blocks and the comprehensive study of their assembly into novel structures, such as alternating polymers or rings, dendrimers or gyroid crystals.

## Author Contributions

R. Mérindol, E. Ducrot and S. Ravaine conceived this research. R. Mérindol and E. Ducrot designed the experimental process and revised the manuscript. R. Khalaf performed most of the experiments. A. Viamonte helped R. Khalaf to perform some of the experiments. R. Khalaf, E. Ducrot and S. Ravaine performed the data analysis. S. Ravaine wrote the manuscript.

## Conflicts of interest

There are no conflicts to declare.

## Acknowledgements

The authors thank S. Yao and I-S. Jo for the functionalization of the copolymer. R. Khalaf thanks the French Ministry of Higher Education, Research and Innovation and Campus France for her PhD grant. This work was supported by the Agence Nationale de la Recherche (POESY project, ANR-18-CE09-0019). We acknowledge funding from IdEx Bordeaux, a program of the French government managed by the Agence Nationale de la Recherche (ANR-10-IDEX-03-02), and from Région Nouvelle-Aquitaine (AAPR 2020-2019-8330510).